\documentstyle[12pt,aaspp]{article}

\begin{document}

\title{The Destruction of a Compact Group of Galaxies}

\author{Rachel A. Pildis}
\affil{Department of Astronomy, University of Michigan, Ann Arbor MI
48109-1090; pildis@astro.lsa.umich.edu}

\begin{abstract}

The very high apparent galaxy densities in Hickson compact groups (HCGs)
should produce copious amounts of galaxy-galaxy interaction, if these groups
are bound dynamical units.  I examine the evidence for such interactions in
HCG 94, a group of seven galaxies with an envelope of diffuse optical light
and an extremely high X-ray luminosity.  Using $ROSAT$ HRI data and deep
three-color optical imaging of the group, I find that both the hot ($k$T
$\sim$ 4 keV) gas and the diffuse optical light trace the same elongated
potential well, which is offset from the galaxy distribution.  The
luminosity and colors of the diffuse optical light imply that the group
galaxies are being ripped apart to form a cD-type remnant galaxy.  The
similarity in shape of the X-ray and optical contours may mean that this
destruction process may require several billion years.

\end{abstract}


\section{Introduction}

Compact groups of galaxies provide a puzzle:  galaxies that are in such
close proximity should merge together rapidly, yet there are many compact
groups and few obvious group remnants such as isolated bright ellipticals.
Some have questioned whether compact groups are real, physical groupings,
but studies of samples such as the Hickson compact groups (HCGs; Hickson
1982) have shown that most group galaxies have accordant redshifts (Hickson
1993) and many of them have morphological signs of interaction (Mendes de
Oliveira \& Hickson 1994).  In addition, recent X-ray investigations
(Ebeling et al.~1994, hereafter EVB; Pildis et al.~1995a, hereafter Paper I)
demonstrate that a significant fraction of elliptical-rich HCGs have
extended emission consistent with hot gas in a group potential (7 of the 8
groups in Paper I with spiral fractions of under 50\% have extended
emission), indicating that they are likely to be gravitationally bound.

In this {\it Letter}, I present new evidence that one X-ray--bright Hickson
compact group---HCG 94---may be near the end of its life as a group,
destroying itself to form a remnant galaxy that may resemble a cD.  HCG 94
is a group with seven members, all but one of them being either elliptical
or S0 (Fig.~1---throughout this paper, I use the galaxy designations of
Hickson [1982,1993] and assume H$_0$=50 km s$^{-1}$ Mpc$^{-1}$).  The
diffuse optical light surrounding the two brightest galaxies has been known
for some time (Hickson 1993), but its extent has not been mapped.  This
diffuse light makes HCG 94 one of the Hickson compact groups most likely
to be close to a merger of its member galaxies.

Serendipitously, this group appeared in a $ROSAT$ PSPC observation of
another group, HCG 93.  No spatial analysis was possible since HCG 94 was
30\arcmin\ off-axis, but spectral analysis gave a temperature of $k$T = $3.7
\pm 0.6$ keV and a luminosity of L(0.07-3.0 keV) = $7.19 \times 10^{43}$ erg
s$^{-1}$ (Paper I---where we assumed a Raymond-Smith thermal model with
absorption due only to the Galactic neutral hydrogen in the direction of
this group and an abundance for the hot gas of 50\% solar.).  This is hotter
and brighter than other X-ray--detected Hickson groups, which have
temperatures of $\sim$1 keV and luminosities of $10^{41-42}$ erg s$^{-1}$
(EVB, Paper I).

The observations to be presented here are a new $ROSAT$ High Resolution
Imager (HRI) observation, important because the galaxies in this group are
too close together to be resolved by the PSPC, and deep three-color ($BVR$)
imaging, obtained as part of a larger survey for low-surface-brightness
optical features in HCGs (Pildis et al.~1995b, hereafter Paper II).  Both
sets of observations reveal that the group potential is offset from the
distribution of galaxies in HCG 94, and that the destruction of the group is
proceeding slowly enough for the stars in the diffuse light to follow the
same potential as the hot gas.

\section{Observations}

The optical observations were obtained in August 1994 with the 2.4m Hiltner
telescope at the Michigan-Dartmouth-M.I.T. Observatory on Kitt Peak,
Arizona.  HCG 94 was imaged in $B$, $V$, and $R$ using a $1024^2$ pixel
back-illuminated Tektronix CCD with a plate scale of
0.275\arcsec\ pixel$^{-1}$.  Calibrations were made using observations of
standard stars from Landolt (1992), and all magnitudes and colors have been
corrected for Galactic extinction.  Details of the reduction procedure for
this and other optical observations of a sample of HCGs will be given in
Paper II.  For HCG 94, the errors in the magnitudes are less than
0.05 mag and the errors in the colors are on the order of 0.05 mag.

The X-ray observations of HCG 94 by the $ROSAT$ HRI were taken in December
1994, with a total live time of 30,045 seconds.  The data were corrected for
vignetting using the PROS package in IRAF.  No signal was found in PHA
channels 1 and 10--16 (as is often found with HRI observations), and thus
those channels were excluded from the analysis.  In order to examine the
spectral hardness of the emission from HCG 94, I created two images:  one
using PHA channels 2-4, which is optimized for temperatures of less than 1
keV, and one using channels 5-9, for higher temperatures.

\section{Results}

Comparison of the two HRI maps reveals strikingly different structures for
the high and low temperature X-ray--emitting gas in this group.  Chi-squared
fits to the low temperature data show that this fainter emission is
circularly symmetric ($Q$=0.81) about galaxy a.  A similar fitting procedure
applied to the high temperature map demonstrates that this bright emission
is best fit by an ellipse of $\epsilon$ = 0.35 ($Q$=0.79) with the major axis
pointing north-south, centered 25\arcsec\ south of galaxy a and almost directly
east of galaxy b.  The centers are determined to within 4\arcsec, so the
displacement between the centers of the two components is significant.

After determining the background level by examining the radial profile of the
emission (the HRI produces very flat images, making this determination
unproblematic), the background was subtracted.  Using a
simplex program, I fit the radial surface brightness profiles of both maps
with the standard hydrostatic-isothermal beta model (Cavaliere \&
Fusco-Femiano 1976):
$$S(r)=S_0\left( 1+\left( {r \over r_{core}}\right)^2\right)^{-3\beta +0.5}$$
convolved with the point-spread function of the HRI.  The
background-subtracted radial surface brightness profiles are shown in Fig.~2,
and the fitting results are listed in Table 1.  This spatial analysis is
very similar to that carried out on PSPC observations of HCGs, as described
in Paper I.  While I used elliptical isophotes to fit a beta model to the
high temperature component, using circular isophotes increases $\chi^2$
but does not change the fitted parameters by more than 1$\sigma$.

The HRI has a very limited capability to provide spectral information.
{}From the relative count rates in the two images (see Table 1) and the
HRI's low sensitivity to high energy X-rays ($k$T $\gtrsim$ 1 keV), one can
determine that the high temperature component dominates, which is consistent
with the 3.7 keV temperature measured with the PSPC.  Since the low
temperature component is centered on galaxy a and has a value of $\beta$
consistent with the values found for elliptical galaxies, it is likely to
also have a temperature typical for hot gas in ellipticals: 0.5--1.0 keV.
The presence of this relatively cool component means that the PSPC
temperature of 3.7 keV is probably an underestimate of the temperature of
the hot intragroup medium (whose $\beta$ value is similar to that of galaxy
clusters).  Thus, as a conservative estimate, I adopt $k$T=4 keV for the
intragroup gas for the remainder of this {\it Letter}.

The diffuse optical emission appears to have the same shape and center as
the high temperature X-ray emission, though its extent is smaller (Fig.~3).
The galaxies in this group are generally to the north and east of the
diffuse light and the hard X-ray emission (compare with Fig.~1).  To a
surface brightness of 26.5 V magnitudes arcsec$^{-2}$, the north-south
extent of the light is 2.4\arcmin (0.17 Mpc) and its ellipticity is
$\epsilon$ = 0.3 (similar to that of the X-ray emission).  Its surface
brightness profile is fairly flat, with the brightest emission located
southeast of galaxy b (22.7 V magnitudes arcsec$^{-2}$) and fainter emission
found north and west of galaxy a (23.6 V magnitudes arcsec$^{-2}$).  The
outer edges of the envelope are somewhat bluer than its central regions, but
the color ranges are small:  \bv\ = 0.95--1.10 and \vr\ = 0.55--0.65.  Its
\bv\ colors are slightly redder than a typical elliptical galaxy (\bv\ =
0.8--1.0), but its \vr\ colors are in the elliptical range of 0.5--0.7
(Gregg 1989).  Galaxies a and b both have blue gradients with increasing
radius, from central colors of \bv\ = 1.20 and \vr\ = 0.75 to colors equal
to those of the inner envelope.  The red \bv\ colors may indicate the
presence of dust in this system.

In order to discriminate between overlapping galaxy light and diffuse light
in the group potential, I ran an ellipse-fitting program on the cores of the
brightest two galaxies, and then forced $r^{1/4}$ profiles with the same
ellipse parameters for their envelopes (details in Paper II).  The two model
galaxies thus produced were then subtracted from the images in all three
filters.  I find that the total luminosity of the diffuse envelope is -23.7
in V, or $2.4 \times 10^{11}$ L$_{\sun}$.  If the two galaxies are included
(M$_{\rm V}$= -22.9 and -22.3, respectively), the system luminosity is -24.3
in V, $4.3 \times 10^{11}$ L$_{\sun}$.

\section{Implications}

If the galaxies in HCG 94 are merging to form an isolated cD-type
elliptical, then the properties of the diffuse X-ray and optical emission
should be comparable to those of cD envelopes.  Schombert (1988) found that
cD envelopes in his large sample have colors of \bv\ = 1.1--1.3 and small to
non-existent color gradients, similar to what is seen in HCG 94.  He also
found a strong correlation between cluster X-ray luminosity and optical
envelope luminosity.  Using that relation, the X-ray luminosity of HCG 94
($7.2 \times 10^{43}$ erg s$^{-1}$) corresponds to a cD envelope with
log(L/L$_{\sun}$)=11.1--11.5.  The measured value for HCG 94,
log(L/L$_{\sun}$)=11.4, is in that range.

Another way to compare the diffuse light in HCG 94 to that in cD envelopes
is to examine the difference between the the total magnitude of the system
and the magnitude of the galaxy if it did not have an envelope.  Using the
models mentioned above for galaxies a and b, this difference is -0.95 if a
and b are together considered to be the central galaxy, and -1.27 if only
galaxy a is used.  Schombert (1986) made histograms of this difference for
both brightest cluster members (BCMs) and cDs, and found values ranging from
0 to -1.7, with average values of roughly -0.6 for BCMs and -1.0 for cDs.
Thus, HCG 94 appears to have an average to bright envelope as compared to
central galaxies of clusters.

HCG 94 is not the sole known example of a system caught in the act of
destroying itself.  Schneider and Gunn (1982) found a similar system that
they considered to be a nascent cD galaxy.  This system, V Zw 311, is a
diffuse halo of optical light surrounding 9 early-type galaxies in the
center of the poor cluster Abell 407.  It has a color of $(g - r)$ = 0.44
(equivalent to a typical elliptical galaxy color of \bv\ = 0.95 [see
Schneider et al.~1983]) and an absolute magnitude of -23.3.  As
in HCG 94, the brightest galaxy in V Zw 311 is a strong radio source and the
X-ray luminosity of Abell 407---$5.8 \times 10^{43}$ ergs s$^{-1}$ (Burns et
al.~1994)---is almost identical to that of this Hickson group.  V Zw 311
differs from the central part of HCG 94 in that there are more galaxies
within its envelope, the system containing it has many more galaxies than an
HCG (its radial velocity dispersion is 590 km s$^{-1}$, while that of HCG 94
is 480 km s$^{-1}$ [Hickson 1993]), and it may be at a more advanced stage
of its evolution since its envelope is symmetric about the galaxy
distribution.  Thus, V Zw 311 may give us an idea of what HCG 94 may look
like in the future as its destruction proceeds and galaxies continue to fall
into it.

The ``fossil group'' discovered by Ponman et al.~(1994) also has some
intriguing similarities with HCG 94.  This apparently isolated elliptical
galaxy has an absolute magnitude of M$_{\rm V}$ = -23.5 and an X-ray
luminosity of $4.5 \times 10^{43}$ erg s$^{-1}$, as well as a rather
red \vr\ color of 0.8.  The outer envelope of the galaxy is even redder,
which differs from the behavior of both ordinary ellipticals and HCG 94.
The fossil group may provide a better comparison to HCG 94 than V Zw 311
since it is not a member of even a poor cluster, but appears to be isolated.

Some have speculated that HCG 94 is not actually an isolated group, but the
core of a poor cluster that is interacting with Abell 2572, a cluster
16\arcmin\ (1.2 Mpc) east at a similar redshift (EVB).  However, isolation
was one of the criteria that Hickson (1982) used to choose his groups, and
the hot gas in HCG 94 is elongated in a north-south direction, showing no
signs that a deep potential to the east is affecting it.  This group may
have an anomalously hot and bright intragroup medium, but there is no {\it a
priori} reason to define it as a cluster core rather than simply a group with
extreme X-ray properties.

Determining what the timescale of the destruction of HCG 94 is could answer
some of the continuing questions about the lifetime of a compact group.  The
median crossing time of HCGs is only 0.016 H$_0^{-1}$ (Hickson et al.~1992),
but simulations, especially those including a common dark matter halo,
demonstrate that compact groups can survive up to $10^9$ years (e.g., Mamon
1987; Barnes 1989; Bode et al.~1993), or even longer if a spectrum of galaxy
masses is used (e.g., Governato et al.~1991).  Even the most optimistic
theories still give merger times that are considerably shorter than a Hubble
time, however.  Such short lifetimes imply that compact group remnants,
which are predicted to be ``dynamically ordinary''---albeit
isolated---ellipticals, should be common (Barnes 1989).  Barnes (1989)
attempted to explain the lack of such remnants by assuming that they are
swept quickly into richer enviroments by the growth of gravitational
clustering, but this would require the somewhat artificial assumption that
the timescales of intragroup and intercluster merging are equal.

In HCG 94, the stars that make up the diffuse light appear to fill the same
potential as the dissipative, X-ray--emitting gas does, implying that the
potential well is not changing on a very rapid time scale.  The crossing
time for stars in the optical envelope is roughly $3\times 10^8$ years (for
velocities on the order of the radial velocity dispersion of HCG 94); to
smooth out the distribution of stars would require a few crossing times, or
over $10^9$ years.  Since the sound-crossing time for 4 keV gas in the same
volume is only $1.6 \times 10^8$ years, the group potential must have been
relatively constant for at least $10^9$ years.  The free-free cooling time
for the hot gas, which has an average electron density of $1.3 \times
10^{-2}$ cm$^{-3}$, is $1.6 \times 10^9$ years, so the assumption that hot
gas has filled this potential for $\sim$10$^9$ years is not unreasonable.
The long timescales seen for HCG 94 imply that HCGs require more time to
merge than many theories predict, perhaps providing a clue to the puzzle
of the abundance of compact groups and the lack of remnants.

\acknowledgements
This paper benefited greatly from illuminating discussions with J. Bregman,
J. Schombert, and A. Evrard.  This material is based upon work supported
under a National Science Foundation Graduate Fellowship and NASA grants
NAGW-2135 and NAG5-1955.

\clearpage

\begin{planotable}{llcccc}
\tablecaption{Fitted spatial properties of X-ray emission}
\tablehead{
\colhead{PHA} & \colhead{center (J2000)} & \colhead{radius} &
\colhead{$r_{core}$} & \colhead{$\beta$} & \colhead{count rate}}
\startdata
2-4 & 23$^h$17$^m$13\fs 1, & 4.2\arcmin\ (0.31 Mpc) & 0.63\arcmin\ $\pm$
0.05\arcmin\ (46 kpc) & 0.56 $\pm$ 0.03 & 0.038 s$^{-1}$\nl
&18\deg 42\arcmin 32\arcsec&&&&\nl
5-9 & 23$^h$17$^m$13\fs 4, & 5.0\arcmin\ (0.36 Mpc) &
0.55\arcmin\ $\pm$ 0.05\arcmin\ (40 kpc) & 0.65 $\pm$ 0.02 & 0.090 s$^{-1}$\nl
&18\deg 42\arcmin 07\arcsec&&&&\nl
\end{planotable}

\begin{figure}

\caption{Grayscale image of HCG 94 in the V band.  North is to the top and
east is to the left, and the image is 3.9\arcmin\ by 4.4\arcmin\ (0.28 by
0.32 Mpc).  The bright galaxy at the center of the image is galaxy a, the
galaxy to the southwest of a is galaxy b, and the remaining five galaxies in
the group (four of which are shown) extend to the northeast of a.}

\caption{Radial profiles of the background-subtracted HRI data for HCG
94. (a) Profile for PHA channels 2-4.  The background level was 0.0105
counts arcsec$^{-2}$.  (b) Profile for PHA channels 5-9.  The background
level was 0.0200 counts arcsec$^{-2}$.}

\caption{V band grayscale image of HCG 94 with HRI X-ray contours overlaid.
The scale is identical to that of Fig.~1.  The X-ray data from PHA
channels 5--9 are smoothed with a gaussian of FWHM=7\arcsec, and the
contours are averaged over the same scale.  The lowest contour is at
0.055 counts per (0.275\arcsec\ $\times$ 0.275\arcsec) pixel, and the
highest is at 0.20 counts per pixel.  Models of the two brightest
galaxies have been subtracted from the optical image, and the stretch was
chosen to maximize the diffuse optical envelope.}

\end{figure}

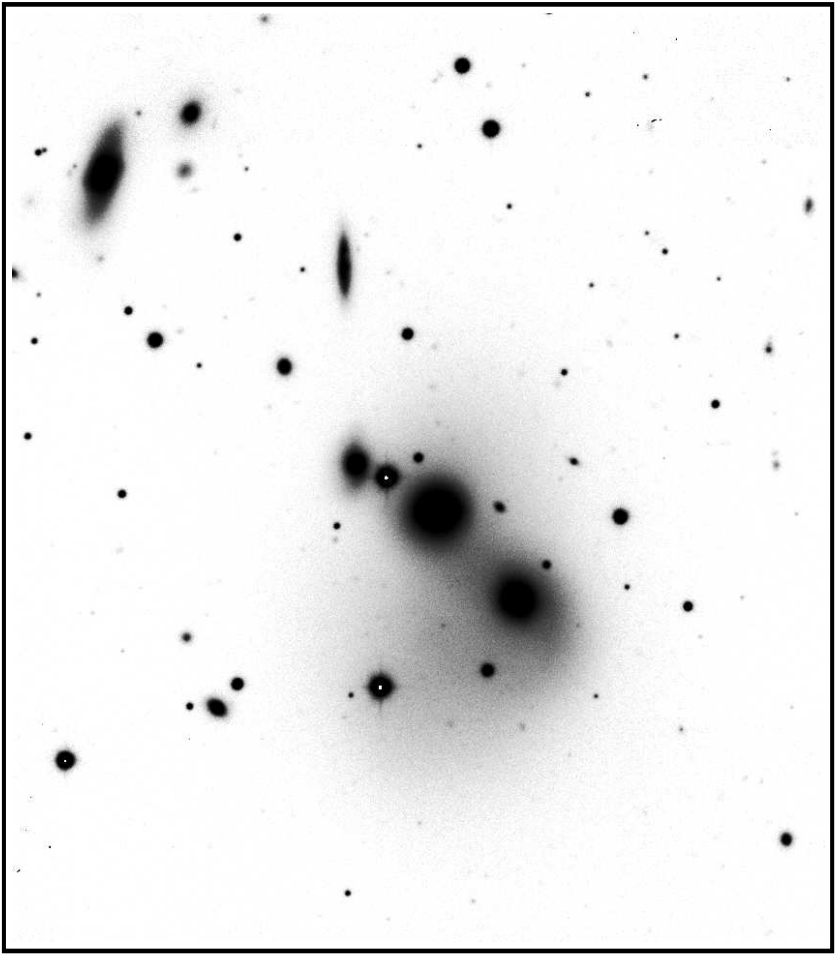
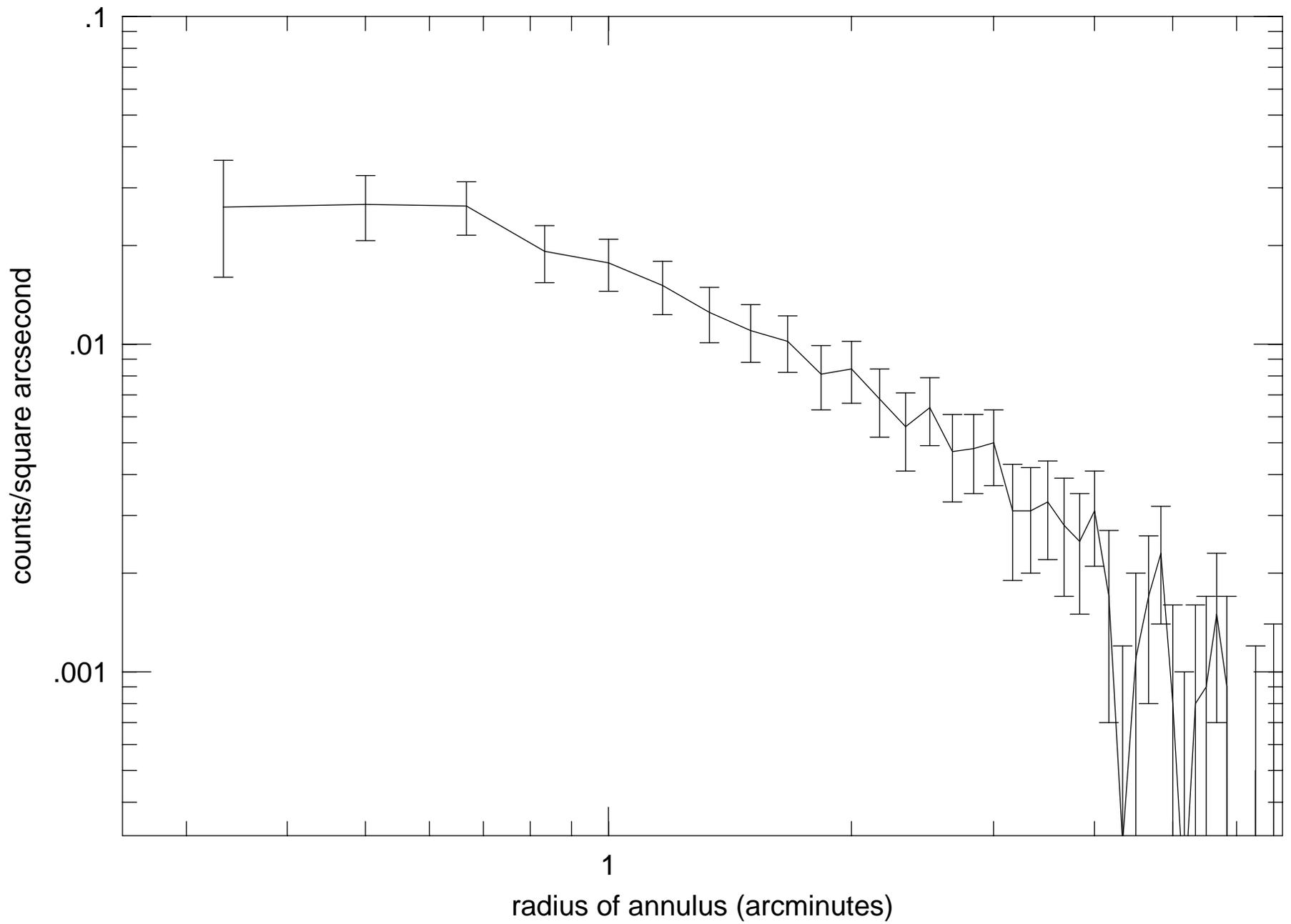
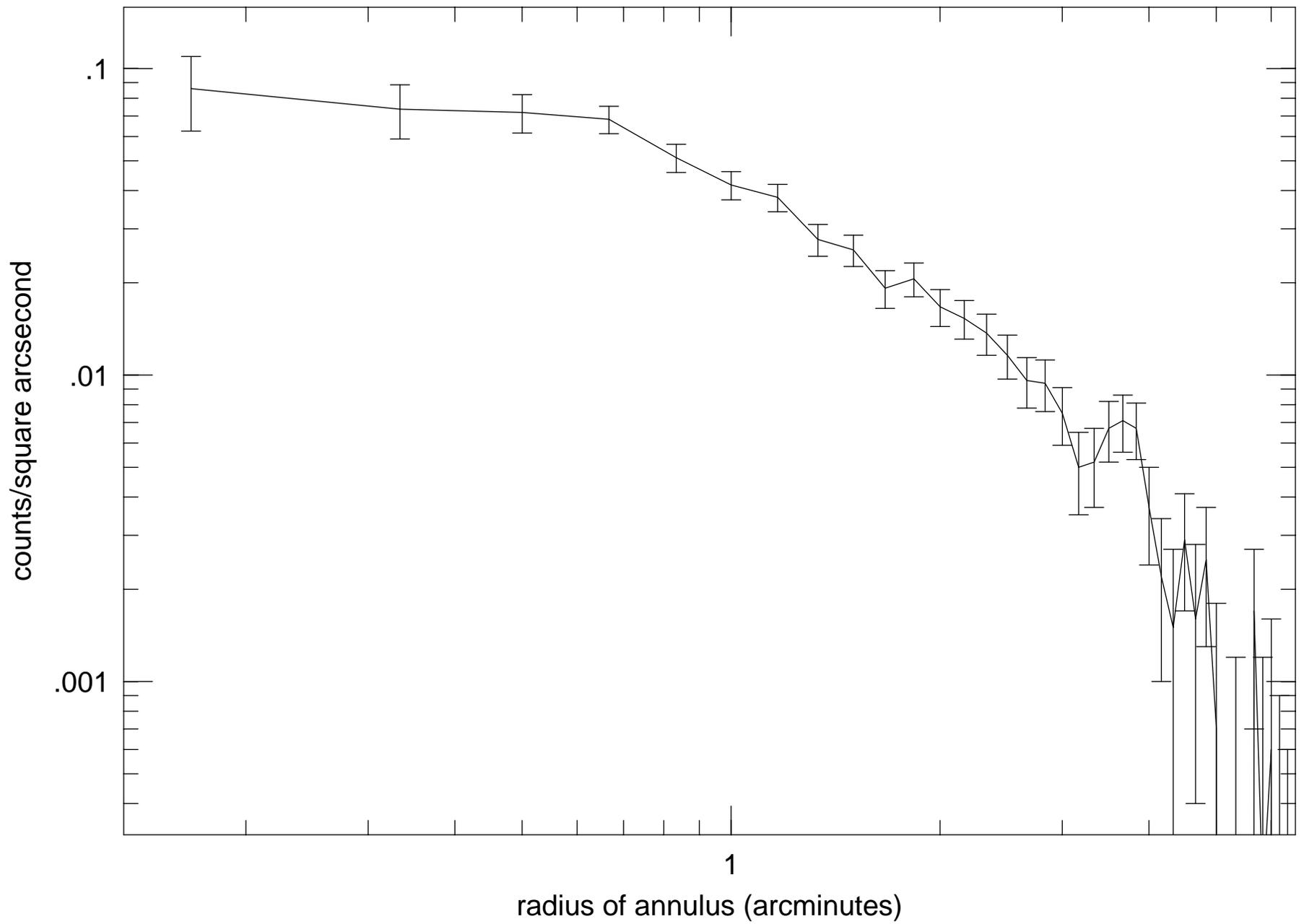
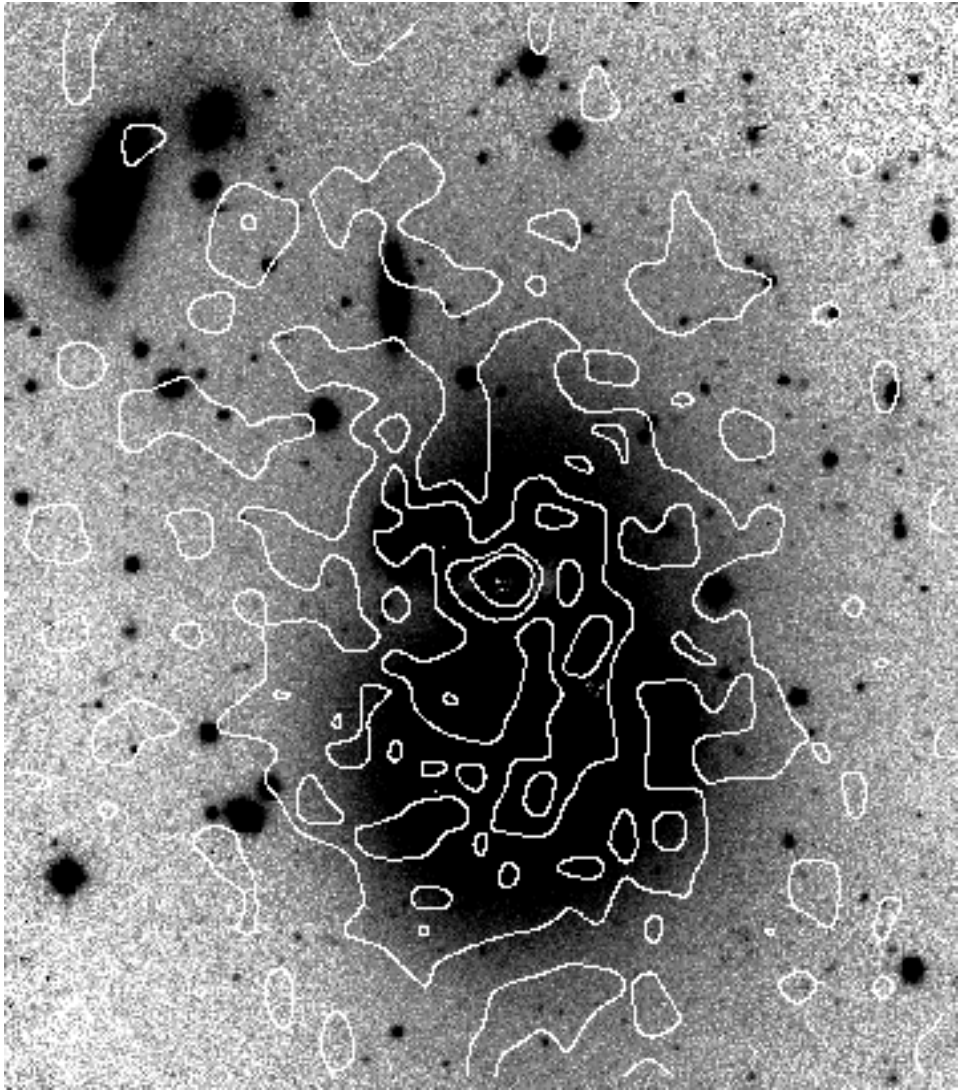
\end{document}